\documentclass[showkeys,showpacs,superscriptaddress,twocolumn,pra]{revtex4-2}
\usepackage{graphicx}
\usepackage{bm}
\usepackage{color}
\usepackage{inputenc}
\usepackage{subfigure}
\usepackage{amssymb}
\usepackage{amsmath}
\usepackage{ifpdf}
\begin{document}

\title{Compressing the atomic cloud in a matter-wave stripe soliton}
\author{Golam Ali Sekh}
\email{skgolamali@gmail.com}
\affiliation{Department of Physics, Kazi Nazrul University, Asansol-713340, India} 
\author{Benoy Talukdar}  
\email{binoyt123@rediffmail.com}
\affiliation{Department of Physics, Visva-Bharati University, Santiniketan-731235, India}

\begin{abstract}
We consider an attractive quasi-one dimensional spin-orbit coupled Bose-Einstein condensate (SOC BEC) confined in a periodic potential produced by the combination of linear and nonlinear optical lattices, and study the effects of squeezing a stripe soliton by varying the  inter-atomic interaction in the nonlinear lattice. It is observed that the nodes on the soliton arising entirely due to the effect of spin-orbit coupling tend to disappear as we increase the squeezing effect to finally get a stable fundamental soliton. This leads us to conclude that external pressure can reduce the effect of spin-orbit coupling in the SOC BEC and even convert the system to a traditional BEC without spin-orbit coupling. We make use of an information theoretic measure to visualize how does the atomic density distribution in the condensate respond to continual reduction in the spin-orbit coupling effect.
\end{abstract}
\maketitle
%
\section*{1. Introduction}
In an atom the nucleus is surrounded by an electron cloud which determines the spectroscopic and chemical properties of the system. It has long been regarded as an interesting curiosity \cite{r1} to examine how do these properties modify if the density distribution of atomic electrons is altered by applying external pressure. For example, as early as 1938 Sommerfeld and Welker \cite{r2} provided a theory in respect of this by confining an atom in a box. But unfortunately, at that time no real physical system could be identified for experimental verification of their claim. In recent years we have, however, plenty of examples regarding quantum confinements and atomic compression. These include atoms in zeolites\cite{r3}, in quantum dots\cite{r4},in metallofullerines \cite{r5} and in many such systems \cite{r6}.In their pioneering work, Sommerfeld and Welker pointed out that the binding energy of an atomic electron  goes on decreasing if the atom is subject to continuously increasing  pressure. Such diminution in binding energy continues until the atom is finally ionized. From the dynamical point of view this observation implies that pressure, which is essentially a thermodynamic variable,can be gainfully employed to reduce the effect of electromagnetic interaction between the electron and nucleus. In the present work we provide an experimentally realizable example in this context. We demonstrate that the effect of spin-orbit coupling in a Bose-Einstein condensate\cite{r7} can be reduced and even eliminated by applying external pressure.  We  work with a one-dimensional attractive SOC BEC loaded in optical lattices\cite{r8} and apply external pressure on it by modulating the atom-atom interaction in the nonlinear lattice.

The experimental realization of SOC BEC \cite{r9,r10} provided a platform to observe supersolidity in the system since the spin degree of freedom is coupled to the density of the condensate \cite{r11,r12}. In fact, after two abortive attempts \cite{r13,r14} to look for supersolidity in solid helium, this intriguing state of matter was first discovered in the SOC BEC \cite{r15,r16} as a stripe soliton. In supersolids, both superfluid and crystalline properties coexist due to simultaneous breaking of phase symmetry and translational invariance\cite{r17,r18}.
    
In section 2 we proceed by assuming that the dynamics of the SOC BEC can be treated within the framework of mean field approximation such that the system is governed  by a set of coupled Gross-Pitaevskii equations \cite{r19}. This provides us an opportunity to reformulate the evolution equations as a variational problem \cite{r20,r21} and use trial  functions and Ritz optimization procedure to obtain the order parameter of the condensate in the supersolid phase which depends rather sensitively on the spin-orbit coupling parameter. The modifications in the density profile corresponding our variational wave function are then studied in section 3 by changing the parameters of the confining potential \cite{r22}. Our primary objective here is to demonstrate how the effect of spin-orbit coupling can be reduced and ultimately made zero by squeezing the SOC BEC using  optical lattices. Since spin-orbit coupling is in the heart of supersolidity, rigging of SOC from a SOC BEC implies a transition from supersolid to superfluid phase. 

 In section 3 we also envisage an information theoretic study to visualize the physical changes that take place in the atomic density distribution in the stripe soliton as we gradually reduce the effect of spin-orbit coupling. In particular, we make use of the well known Fisher information\cite{r23} for our analysis. An appealing feature of this information measure is its local character as opposed to the global nature of other such functionals like the variance, Shannon, Tsallis and Reneyi entropies \cite{r24}. The local character of Fisher information provides an enhanced sensitivity to strong changes even over a small-sized region in the density functional.  For a position-space normalized to unity probability density $\rho(x)$, the position-space  Fisher information is given by
\begin{eqnarray} 
I_{\rho}=\int_{-\infty}^{+\infty}\rho(x)\left[\frac{d}{dx}\ln\rho(x)\right]^2dx
\label{eq1}
\end{eqnarray}
with the  corresponding momentum-space results 
\begin{eqnarray} 
I_{\gamma}=\int_{-\infty}^{+\infty}\gamma(p)\left[\frac{d}{dp}\ln\gamma(p)\right]^2dp,
\label{eq2}
\end{eqnarray}
where $\gamma(p)$  represents the momentum- or $p$-space probability density. Although Fisher information was originally introduced as a measure of intrinsic accuracy in statistical estimation theory, it was given quantum mechanical generalization in early 1970's \cite{r25,r26}. The quantum extension was used to derive the so-called Cramer-Rao inequalities \cite{r27, r28} and an information theoretic uncertainty relation\cite{r29}, which for one-dimensional systems reads as
\begin{equation}
I_\rho I_\gamma \geq 4.
\label{eq3}
\end{equation}
We compute numbers for $I_{\gamma}$  and  $I_{\rho}$  as a function of the lattice parameters with a view to visualize the physical change that takes place in the  atomic density distribution of the condensate  as we gradually reduce the effect of the spin-orbit coupling by applying external pressure. Finally, in section 4 we summarize our outlook on the present work and make some concluding remarks.

\section*{2.	Spin-orbit coupled Bose-Einstein condensate in optical lattices}  
 In the mean field approximation, the dynamics of the quasi-one-dimensional SOC BEC having equal contributions from Rashba and Dresselhaus couplings is governed by the Gross-Pitaevskii equation (GPE) \cite{r9,r10} 
\begin{eqnarray}
i \partial_t {\psi}_j\!&=&\left(\!-\frac{1}{2}\partial_x^2+\!i(-1)^j\! k_s \partial_x\!+\!V(x)\!+\!\gamma(x) |{\psi}_{j}|^2\right.\nonumber\\
&+&\left.\beta(x) |{\psi}_{3-j}|^2\right){\psi}_j+\Omega {\psi}_{3-j}, \hspace{0.75cm} j=1,2.
\label{eq4}
\end{eqnarray}
Here  ${\psi}_1, {\psi}_2 $ stand for the pseudo-spin components of the condensates's ordered parameter corresponding to the hyperfine states labeled by $|\uparrow\rangle\equiv |1,\,0\rangle$ and $|\downarrow\rangle \equiv |1,\,-1\rangle$, and $V(x)$ is the external trapping potential. The parameters $\gamma$ and $\beta$  represent intra- and inter-atomic interactions. In writing Eq. (\ref{eq4}) we have introduced  scaled quantities  $x\rightarrow x/\sqrt{\hbar/m\omega_\perp}$, ${\psi}_j \rightarrow {\psi}_j \sqrt[4]{\hbar/m\omega_\perp}$ and $t\rightarrow t\,\omega_\perp$ in term of the transverse trapping frequency $\omega_\perp$.  The wave number of the Raman laser or strength of the spin-orbit coupling as well as the Rabi frequency are also scaled as  $k_s\rightarrow k_s \sqrt{\hbar/m\omega_\perp} $ and $\Omega \rightarrow \Omega/\omega_\perp$ respectively. 

For attractive inter-atomic interaction, Eq.(\ref{eq4}) can support different types of soliton solutions \cite{r19} including the so-called stripe soliton \cite{r12,r15,r16,r30} characterized by a modulated density profile originating  from  the effect of synthetic spin-orbit coupling  in the BEC. Since our objective here is to examine the effect of pressure on the condensate by loading it in optical lattices, we assume the following
\begin{subequations}
\begin{eqnarray}
V(x)&=&V_0 \cos(k_L x),\\
\gamma(x) &=&\gamma_0+\gamma_1\,\cos(k_N x)\\
{\rm and}\hspace{1cm}\beta(x) &=& \beta_0+\beta_1\,\cos(k_N x)
\end{eqnarray}
\label{eq5}
\end{subequations}
with $k_L$ and $k_N$ are the wave numbers of the linear and nonlinear optical lattices. Relatively recently, Wong et. al. \cite{r31} showed that spatially  modulated nonlinearity can protect the stability of vortex line structure in a two dimensional SOC BEC. But, we shall make use of parameters in Eq.(\ref{eq5}) to provide a useful control over the effects  spin-orbit coupling in a stripe soliton supported  by an attractive quasi-one dimensional SOC BEC.

Equation (\ref{eq4}) is not exactly solvable. Consequently, we consider a variational formulation of the problem and find a Lagrangian density which via Legendre transformation leads to a Hamiltonian density such that the energy functional of the system is given by
\begin{eqnarray}
E[\psi_j]&=&\int dx [\sum_{j=1}^2 (\frac{1}{2} |{\partial_x \psi_j}|^2+\frac{1}{2} \gamma (x) |\psi_j|^4\nonumber\\
&+& (-1)^j\frac{i k_s}{2}(\psi^*_j\psi_{jx}-\psi_j\psi^*_{jx}) +V(x) |\psi_j|^2 )\nonumber\\
&+&\Omega(\psi_2\psi_1^*-\psi_1\psi_2^*) +\beta (x) |\psi_1|^2 |\psi_2|^2].
\label{eq6}
\end{eqnarray}
We introduce  the trial function of our system as  
\begin{eqnarray}
\psi_j(x)&=& A_j\, {\rm sech}(\frac{x}{a})\,\exp[(-1)^j i\, k_j x+ i\phi_j]\,.
\label{eq7}
\end{eqnarray}
with parameters $A_j$, $k_j$ and $a$. In the absence of optical lattices i. e. when   $V_0=\gamma_1=\beta_1=0$, Eq.(\ref{eq4}) can be solved exactly by the use of multi-scale perturbation theory\cite{r19}. In the presence of optical lattices the  parameters $A_j$, $k_j$ and $a$ are allowed to vary with a view to catch the effects of compression in the order parameter through energy optimization procedure. The wave function $\psi_j(x)$ are normalized  such that $N=\int_{-\infty}^{+\infty}\,[|\psi_1|^2+|\psi_2|^2] \,dx=N_1+N_2$ (say).  From Eqs. (\ref{eq6}) and (\ref{eq7}) the effective energy of the system is obtained as
\begin{eqnarray}
E_0&=&\sum_{j=1}^2\left[\frac{A_j^2  }{3 a}+a k_j^2 A_j^2+2a k_j k_s A_j^2+\frac{2}{3}a\gamma_0 A_j^4\right.\nonumber\\
&+& 2 \pi  a^2 A_j^2 \frac{k \,V_0}{{\rm sinh}[\pi k a]}+\frac{2\pi}{3}a^2 k  (1+a^2 k^2)A_j^4 \gamma_1\ \nonumber\\
&\times& \!\!\left.\frac{1}{ {\rm sinh}[\pi k a]} \!+\!2\pi \Omega a^2 A_1 A_2  \frac{k_j \cos\phi}{{\rm sinh}[\pi k_j a]} \right]\nonumber\\&+&
\frac{4\pi \beta_1 }{3 {\rm sin}[\pi k a]}k a^2 A_1^2 A_2^2(1\!+a^2 k^2)
 +\frac{4}{3}\beta_0 a A_1^2 A_2^2.
\label{eq8}
\end{eqnarray}
Here  $\phi=\phi_1-\phi_2$, $k=k_L=k_N$. The phase difference $\phi= n\pi$ with $n=0,1,2,3$. We calculate values of the parameters of the trial solution by minimizing the energy functional in Eq. (\ref{eq8}).
\begin{figure}[h!]
\begin{center}
\includegraphics[width=0.4\textwidth]{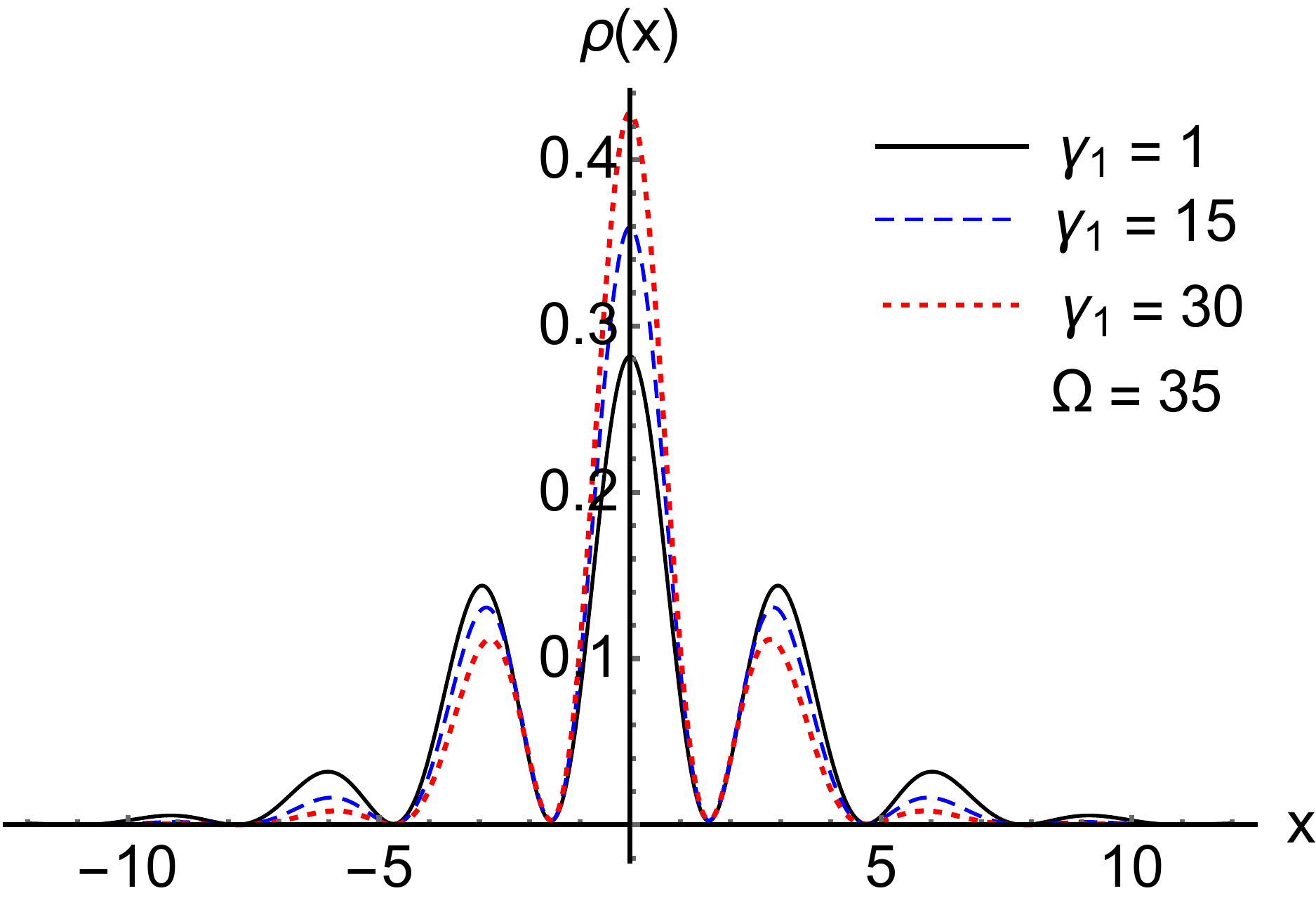}
\vskip -0.1cm
\includegraphics[width=0.4\textwidth]{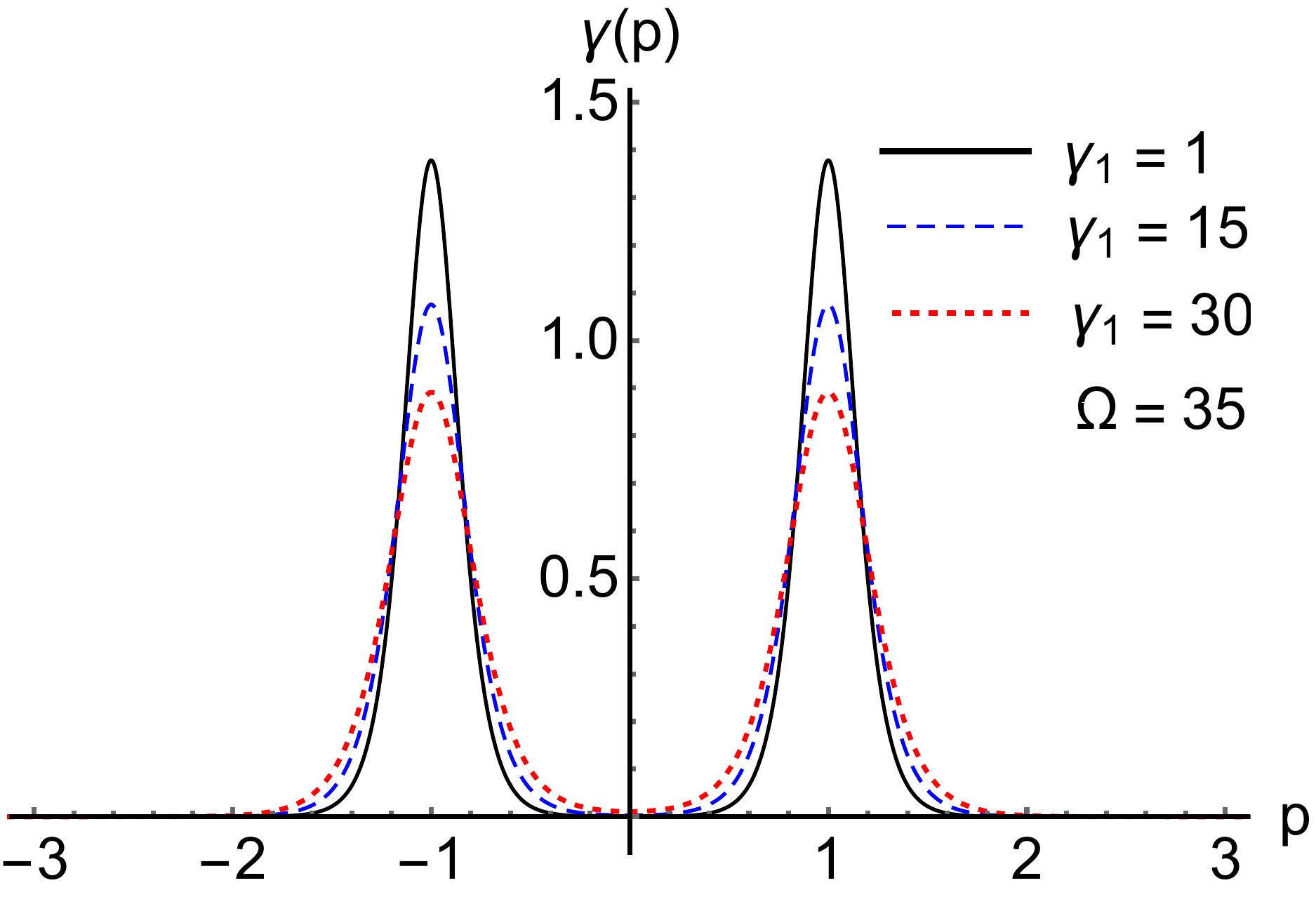}
\vskip 0.01cm
\caption{Both coordinate- and momentum -space density profiles of spin-orbit coupled BECs with $\Omega=35$, $N=1(N_1=0.6,N_2=0.4)$ but different values of nonlinear lattice parameter. Here we have used $k_s=8$(spin-orbit coupling), $k=0.75$,$\gamma_0=\beta_0=-1$,$V_0=-3$. Red, black and blue curves represent respectively density distributions for NOL strengths $1$, $15$ and $30$.}
\label{fig1}
\end{center}
\end{figure}

\section*{3.	 Density profile of the stripe soliton as a function of lattice parameters}
In the absence of optical lattices, the choice for the values of $k_s$ and $\Omega$  allows one to distinguish two different regions \cite{r19} in the linear energy spectrum of the system in Eq.(\ref{eq4}). Region 1 is characterized by $k_s^2<\Omega$  and the dispersion curve has a single minimum such that the associated GPE supports a usual $sech$ soliton solution. But in region 2, with $k_s^2>\Omega$, the dispersion curve possesses two minima at momenta, {say, $\pm k_1$} of the system. We can have two different solutions of the GPE corresponding to these minima. In addition, we can have a linear superposition of these solutions that form a stripe phase \cite{r9,r30}. We have found the density profile for the so-called stripe soliton in the form

\begin{eqnarray}
\!\rho(x)\!=\!C_{0x}\!\left(A_1^2\!+\!A_2^2\!+\!2 A_1 A_2 \!\cos (2 k_1 x)\right)\!\text{sech}^2\!\left(\frac{x}{a}\right)\!\!.
\label{eq9} 
\end{eqnarray}
Here $C_{0x}$ is a constant related to the normalization $\int_{-\infty}^{+\infty} \rho(x)\, dx=1$. We have verified that in the appropriate limit Eq.(\ref{eq9}) gives the result for the density distribution calculated in the absence of the confining potential \cite{r19}. The momentum-space probability corresponding to the position-space result in Eq.(\ref{eq9}) is given by
\begin{equation}
\gamma(p)=C_{0p}(C_1^2|\Gamma_1(p)|^2+C_2^2|\Gamma_2(p)|^2),
\label{eq10}
\end{equation}
where $C_1=A_1+A_2$ and $C_2=A_1-A_2$. The factor $C_{0p}$  is the momentum-space analog of $C_{0x}$. Also we have
\begin{eqnarray}
\Gamma_1(p)&=&	\frac{a \sqrt{\pi } }{2\sqrt{2}} \text{sech}\left(\frac{1}{2} \pi  a (k_1+p)\right)\nonumber\\&+&\frac{a \sqrt{\pi }}{2 \sqrt{2}D}\left( \cosh \left(\frac{1}{2} \pi  a (k_1+p)\right)\right.\nonumber\\&+&\left. \sinh \left(\frac{1}{2} \pi  a_1 (k_1+p)\right)\right)
\label{eq11}
\end{eqnarray}
and 
\begin{eqnarray}
\Gamma_{2}(p)&=& \frac{-i a \sqrt{\pi } }{2\sqrt{2}} \text{sech}\left(\frac{1}{2} \pi  a (k_1+p)\right)\nonumber\\&+&\frac{i a\sqrt{\pi }}{2 \sqrt{2}D}\left( \cosh \left(\frac{1}{2} \pi  a (k_1+p)\right)\right.\nonumber\\&+&\left. \sinh \left(\frac{1}{2} \pi  a (k_1+p)\right)\right).
\label{eq12}
\end{eqnarray}
with $D=\sinh (\pi  a k_1)+\cosh (\pi  a k_1)+\sinh (\pi  a p)+\cosh (\pi  a p)$.
\begin{figure}[h!]
\begin{center}
\includegraphics[width=0.4\textwidth]{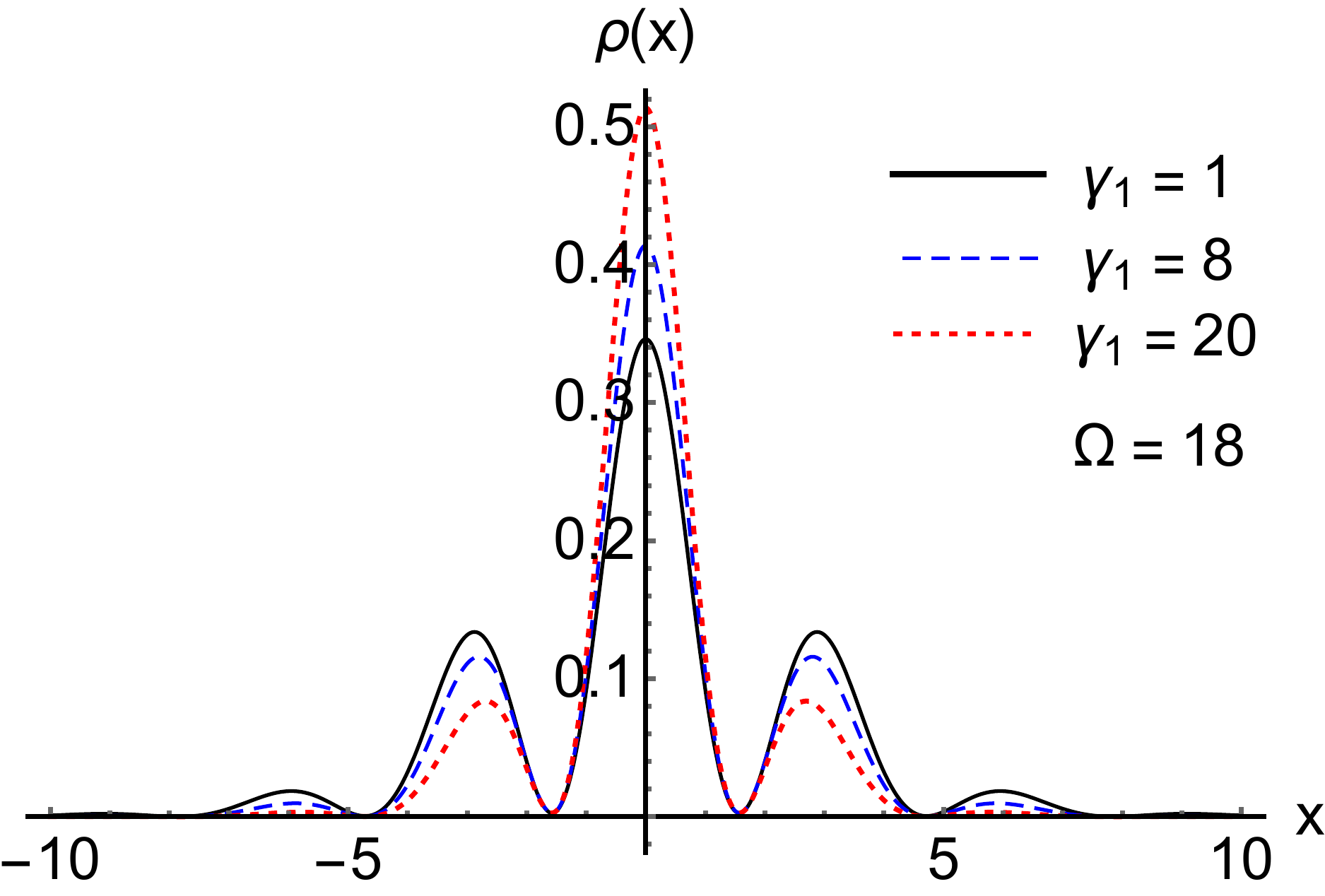}
\vskip -0.1cm
\includegraphics[width=0.4\textwidth]{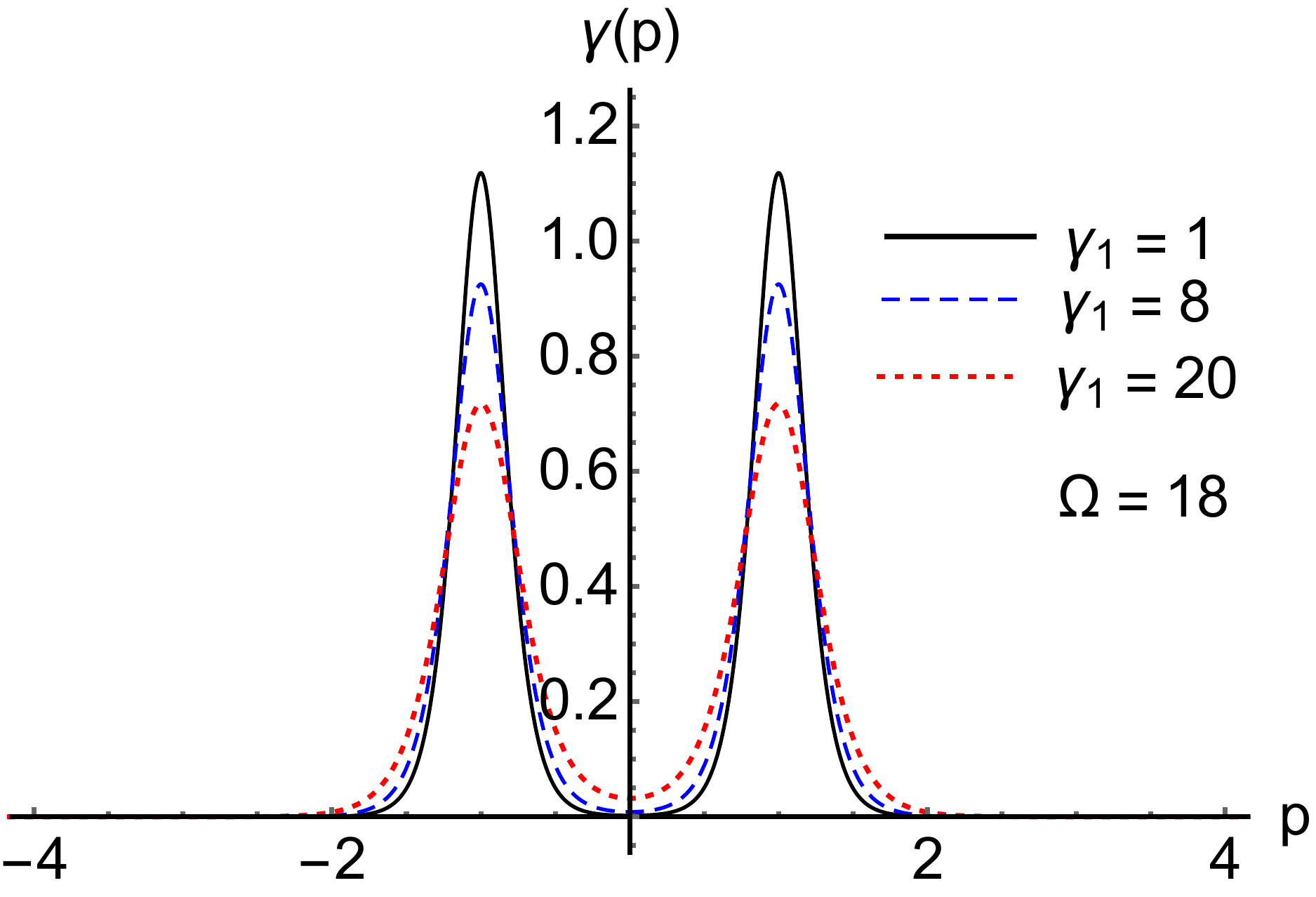}
\vskip 0.1cm
\caption{Density plots (coordinate and momentum spaces) for $\Omega=18$  and different values of nonlinear lattice parameter. Solid, dashed and dotted lines give the density plots for $\gamma_1=1,\, 8$ and $20$ respectively.}
\label{fig2}
\end{center}
\end{figure}
\begin{figure}[h!]
\begin{center}
\includegraphics[width=0.4\textwidth]{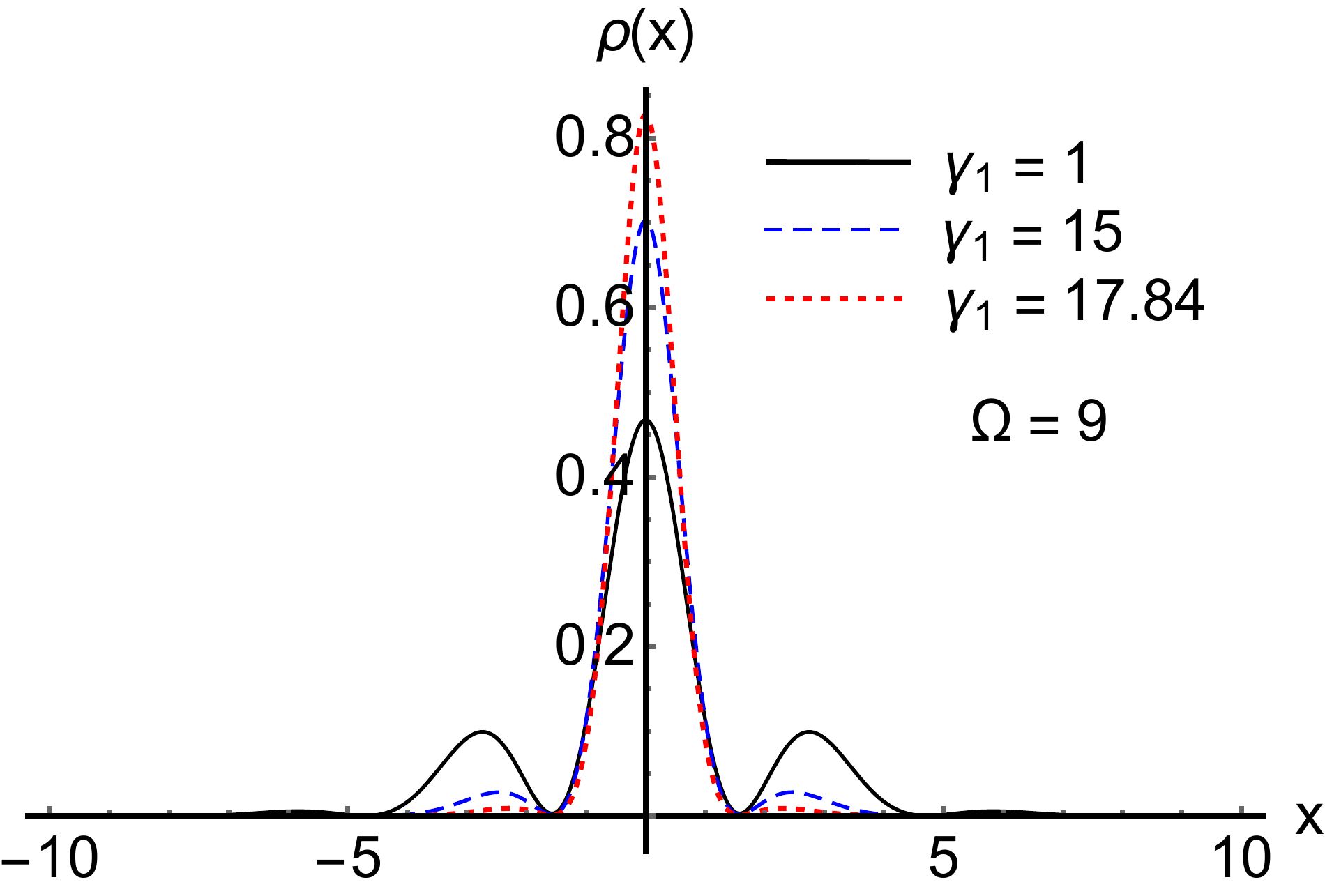}
\vskip -0.1cm
\includegraphics[width=0.4\textwidth]{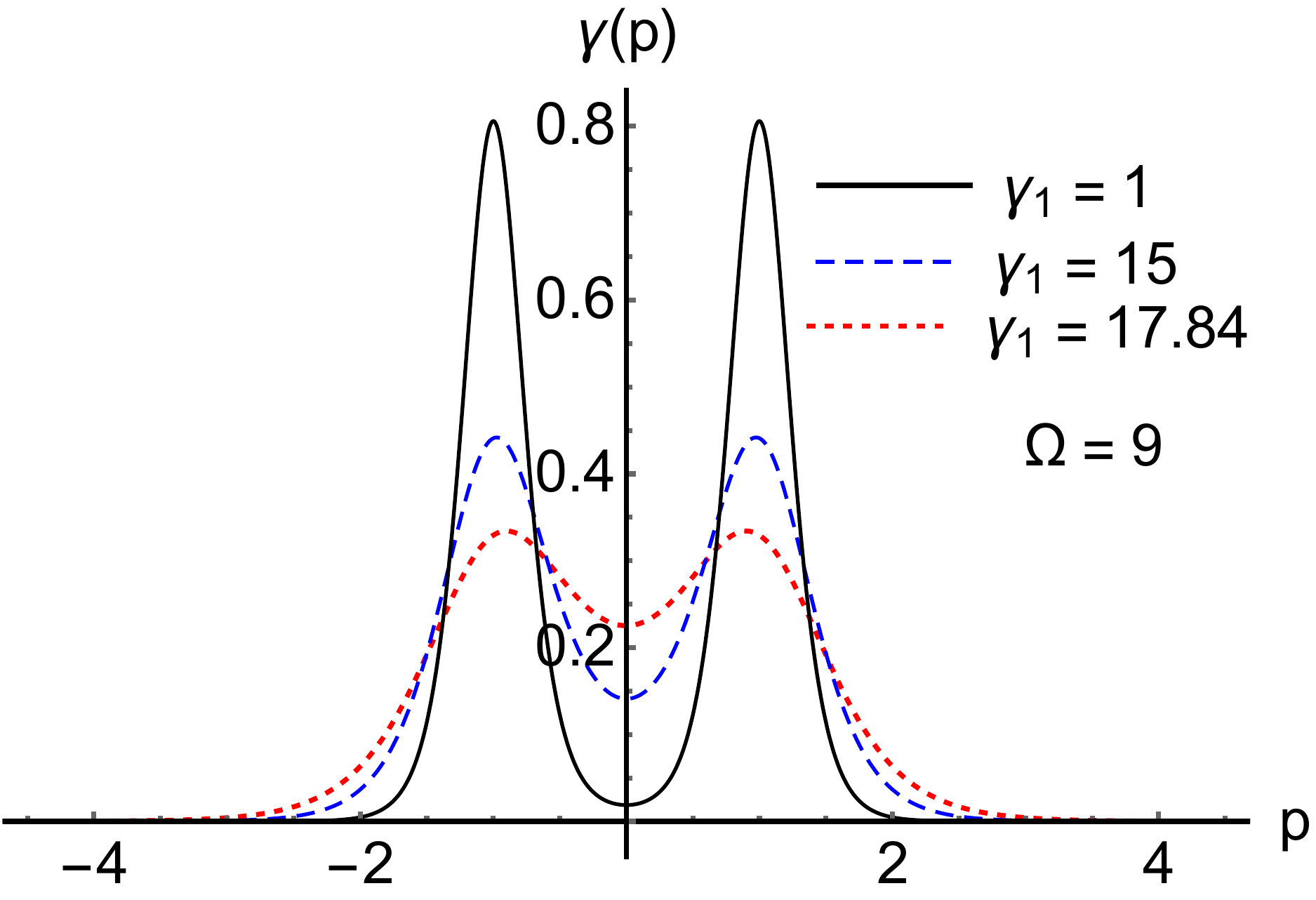}
\vskip 0.1cm
\caption{Density distributions in both coordinate and momentum spaces  of spin-orbit coupled BECs  with $\Omega=9$ for different values of nonlinear lattice.  Other parameters  are kept same with those used in Fig. \ref{fig1}. Red, black and blue curves represent respectively density distributions for  $\gamma_1=\,1$, $15$ and $17.84$.}
\label{fig3}
\end{center}
\end{figure}

We now make use of Eqs. (\ref{eq9}) and (\ref{eq10}) to visualize the effect of compressing the modulated density profile of the stripe soliton by varying the parameters of the optical lattices and, thereby, try to attain some added realism for the interplay between spin-orbit coupling and external pressure on the condensate. In Fig. \ref{fig1} (upper plot) we display $\rho(x)$  as a function of $x$ for three different values of the nonlinear lattice parameter $\gamma_1$. Here we have chosen to work with $k_s=8$ and $\Omega=35$. Each curve for $\rho(x)$  having different values of $\gamma_1$ is characterized by a central maximum with relatively weak undulations on either side of it. The solid curve portrays the variation of the density distribution when the effect of the nonlinear lattice is minimum ($\gamma_1=1$).  The dashed curve represents a similar variation in the density profile for $\gamma_1=15$ . Looking closely onto these curves (solid and dashed) we see that as we increase strength of the nonlinear lattice the central peak becomes stronger or augmented while the peaks on the either side of it, having their dynamical to the spin-orbit coupling, become weaker. The dotted curve gives the plot of $\rho(x)$ for still a higher value of $\gamma_1$ ($\gamma_1=30$ ). Here strengthening of the central peak and weakening of the  associated  secondary maxima are seen to be more pronounced. The corresponding curves for momentum-space density distribution  $\rho(p)$  as a function of $p$ are displayed in the lower plot. All these curves corresponding to $\gamma_1=1,15$  and $30$  respectively, have two peaks lying symmetrically about the $\gamma(p)$ axis. The dashed and dotted curves show that the height of each peak reduces as we go to higher value of  $\gamma_1$ and at the same time the central minimum tends to rise up.

To visualize the effect of pressure on SOC BEC more clearly it may be interesting to work with some lower values of the Rabi frequency, say $\Omega=18$ and $9$ for which we shall have fewer number of nodes \cite{r7} compared to those appearing in Figure \ref{fig1}.

\begin{figure}[h!]
\begin{center}
\includegraphics[width=0.4\textwidth]{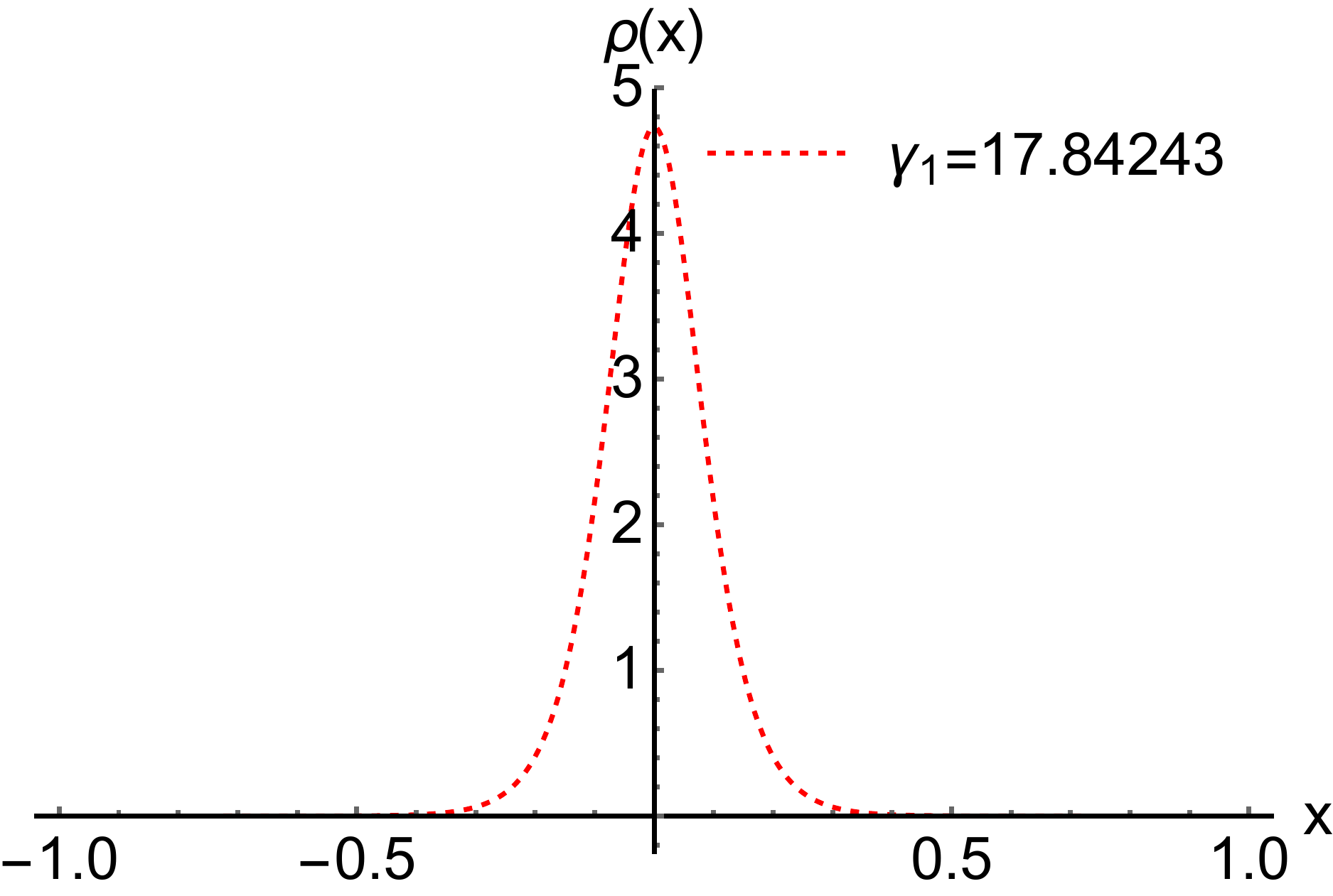}
\vskip -0.1cm
\includegraphics[width=0.4\textwidth]{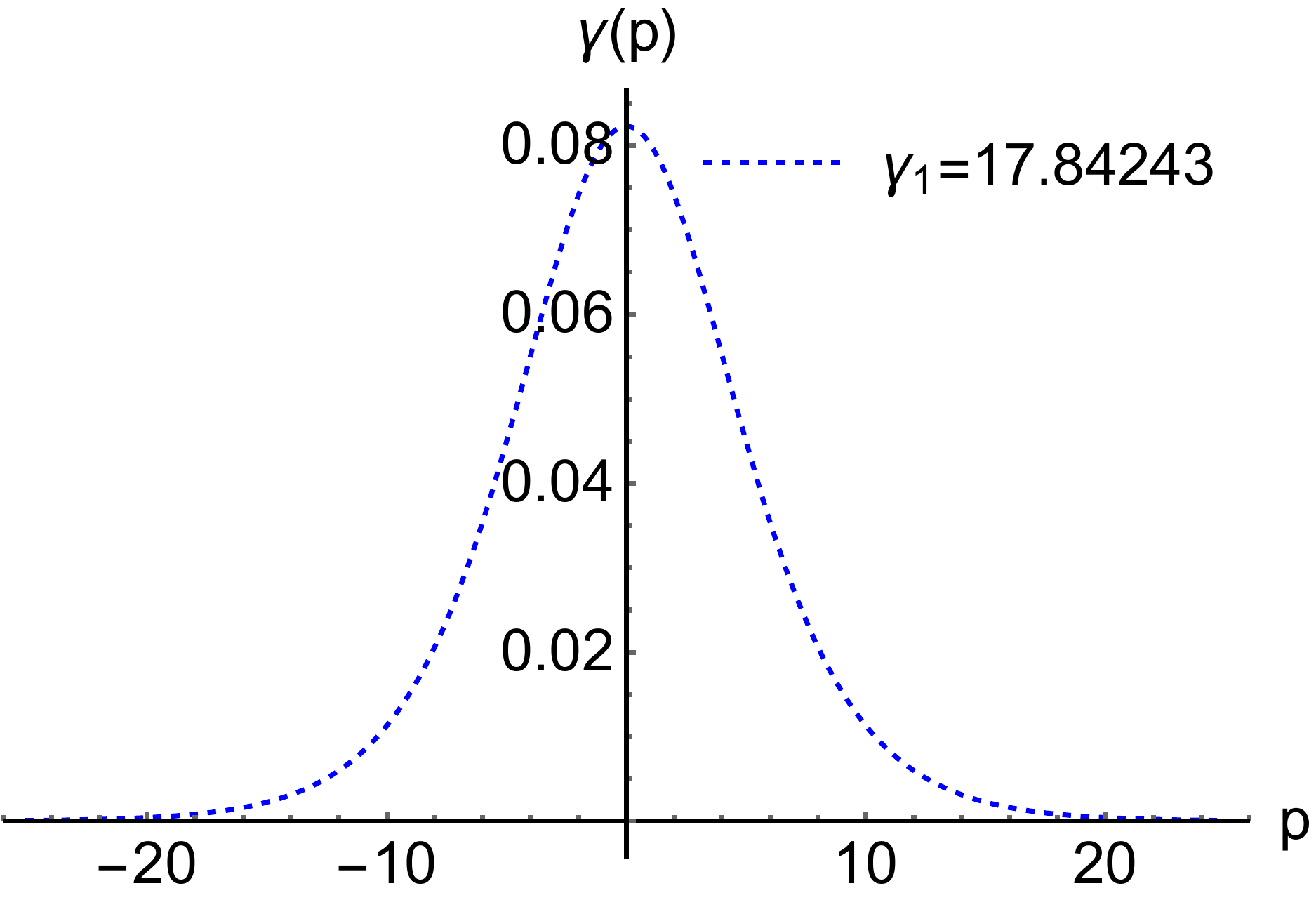}
\vskip 0.1cm
\caption{Coordinate - and momentum -space density plots  for $\gamma_1=17.84243$  and  $\Omega=9$.}
\label{fig4}
\end{center}
\end{figure}

In Figs. \ref{fig2} and \ref{fig3} we present plots of  $\rho(x)$  and $\gamma(p)$ for  $\Omega=18$ and $9$   respectively with   $k_s=8$ . In the rest of the paper we shall make use of the same value of $k_s$  wherever it appears. {In Fig. \ref{fig2} the values of $\gamma_1$  for the solid, dashed and dotted lines are $1$, $8$  and $20$  respectively}. The curves in these plots confirm that response of density profile, $\rho(x)$ or $\gamma(p)$, to compression  is similar to that observed in Figure \ref{fig1}. But here the effect of compression appears to be more pronounced. For example, the peak of $\rho(x)$  at $x=0$ as shown by the dotted curve ($\gamma_1=20$ ) is larger than the height of the central peak of  $\rho(x)$ in Figure 1 ($\gamma_1=30$ ) with significantly more depression of the nodes on the wings. The behavior of the dashed and dotted curves in the plot for $\gamma(p)$ (Fig.\ref{fig2}) is similar to that observed in right plot of Figure \ref{fig1}. But significantly enough, the elevation of the minimum of the dotted curve now appears to be appreciable.

The plots in Fig. \ref{fig3} tend to provide a rather confirmatory evidence to realize how compression of the atomic cloud really reduces the effect of spin-orbit coupling in a Bose-Einstein condensate. Looking closely into the dotted curve for $\rho(x)$ ($\gamma=17.84$ ) we see that it has practically no nodes on the wings verifying about complete removal of the spin-orbit coupling effect due to substantial amount of external pressure on the system. The minimum of the corresponding curve for $\gamma(\rho)$ is now $>0.2$. Finally, as shown in Fig. \ref{fig4}, we have found that for $\gamma_1=17.84243$  and  $\Omega=9$  the external pressure  converts the modulated density profiles of the supersolid phase of the BEC  to conventional solitonic density profiles that is free from the effect spin-orbit coupling.

Based on the expressions for the density profiles in Eqs.(\ref{eq9}) and (\ref{eq10}) we now present numbers for $I_\rho$, $I_\gamma$  and the corresponding information-based uncertainty $I_\rho I_\gamma$ in order to realize how the distribution of cold atoms inside the condensate changes as we gradually reduce the effect of spin-orbit coupling by increasing the value of $\gamma_1$. We display, in Table I, the results for these information theoretic measures for $\Omega=35$, $18$ and $9$  computed by the use of Eqs. (\ref{eq1}) and (\ref{eq2}). The numbers in the table have been generated by varying the parameter $\gamma_1$  rather arbitrarily. 

It is well known that larger values of position-space Shannon entropy are associated delocalization in the density distribution while smaller values with localization\cite{r32}. As opposed to this $I_\rho$ provides a measure of localization \cite{r7,r33}; smaller (larger) values relate to delocalization (localization). The numbers for $I_\rho$ in Table I indicate that, irrespective of the values of $\Omega$, the result position-space Fisher information is an increasing function of $\gamma_1$. For $\gamma_1=1$ the value of $I_\rho$ is minimum. This indicates that the atomic density distribution in the stripe soliton is highly delocalized when the  effect of spin-orbit coupling is very large. Moreover, rise in the values of $I_\rho$ for higher values of $\gamma_1$   confirms that the density distribution gradually becomes localized as the spin-orbit coupling effect is reduced by exerting pressure on the condensate. As expected the results for the momentum-space Fisher information $I_\gamma$  decrease as $\gamma_1$  takes up larger values. The values of the uncertainty product  for every value of $\Omega$  is a decreasing function of $\gamma_1$. Thus one can infer that the quantum mechanical aspect of the stripe soliton is reduced under external pressure.

\begin{widetext}
\begin{center}
\begin{table}
\begin{tabular}{cccccccccccc}
\hline
& $\Omega=35$& & & &$\Omega=18$&  & & &$\Omega=9$ & & \\
\hline
\hline
$\gamma_1$&$I_\rho$&$I_\gamma$&$I_\rho I_\gamma$&$\gamma_1$&$I\rho$&$I_\gamma$&$I_\rho I_\gamma$&$\gamma_1$&$I\rho$&$I_\gamma$&$I_\rho I_\gamma$\\
\hline
1&3.3080&40.4958 &133.9600&1&3.3625&26.5891&89.4060&1&3.4984&13.4015&46.8843\\
6&3.3320&33.0624&110.1560&4&3.3923&22.2068&75.3331&3&3.5463&11.1061&39.3846\\         
 11&3.3561&27.7605&93.1632&7&3.4243&18.9186&64.7413&5&3.5980&9.2375&33.2361\\
16&3.3802&23.8297&90.5482&10&3.4531&16.3203&56.3555&7&3.6551&7.6662&28.0205\\
21&3.4045&20.7707&70.7135&13&3.4848&14.1895&49.4472&9&3.7203&6.3126&23.4850\\
26&3.4292&18.2925&62.7277&16&3.5180&12.3878&43.5797&11&3.7980&5.1230&19.4572\\ 
31&3.4544&16.2206&56.0329&19&3.5532&10.8286&38.4763&15&4.0336&3.0689&12.2788\\ 
36&3.4806&14.4451&50.2774&22&3.5912&9.4542&33.9520&17&4.2841&2.0674&8.8570\\
\hline 
\end{tabular}
\caption{Variation  of Fisher information and corresponding uncertainty  with  increasing values of $\gamma_1$ (diminution in spin-orbit coupling effect) for $\Omega=35$, $18$ and  $9$  ($k_s=8$ and  $V
_0=-3$).}
\end{table}
\end{center}
\end{widetext}

\section*{4. Concluding remarks}
It is true that Sommerfeld and Welker \cite{r2} had difficulty to find experimental evidences in support of their theory which tells us that the binding energy of an electron in the atom decreases gradually as it is subjected to continuously increasing pressure. Admittedly, the predicted diminution in binding energy implies that the pressure applied on the atom affects electromagnetic interaction between the electron and nucleus. Thus we found it interesting to demonstrate if the effect of spin-orbit coupling could be removed from an SOC-BEC. The number of nodes on the stripe soliton was found to decrease gradually as we increase the pressure on it and ultimately we arrive at a nodeless stable soliton. We also verified that by reducing the effect spin-orbit coupling from the stripe phase of the BEC we
go from a de-localized to localized atomic density distribution. We note that studies in the interplay between optical lattices and BEC are of utmost interest in the physics of cold atoms \cite{r34}. Working along this line of investigation we provide here an experimentally realizable example in support of the conjecture made by Sommerfeld and Welker.

\subsection*{Acknowledgment} One of the authors (GS) would like to  acknowledge the funding from the ``Science  and Engineering Research Board, Govt.of India" through Grant No. CRG/2019/000737.


\end{document}